# Fuzzy inference system application for oil-water flow patterns identification

WU Yuyan[1,2]*, GUO Haimin[1,2], SONG Hongwei[1,2], DENG Rui[1,2]

*(1. College Physics and Petroleum Resources, Yangtze University, Wuhan 430100, China; 2. Key Laboratory of Exploration Technologies for Oil and Gas Resources, Ministry of Education，Yangtze University, Wuhan 430100, China)*

**Abstract.** With the continuous development of the petroleum industry, long-distance transportation of oil and gas has been the norm. Due to gravity differentiation in horizontal wells and highly deviated wells (non-vertical wells), the water phase at the bottom of the pipeline will cause scaling and corrosion in the pipeline. Scaling and corrosion will make the transportation process difficult, and transportation costs will be considerably increased. Therefore, the study of the oil-water two-phase flow pattern is of great importance to oil production. In this paper, a fuzzy inference system is used to predict the flow pattern of the fluid, get the prediction result, and compares it with the prediction result of the BP neural network. From the comparison of the results, we found that the prediction results of the fuzzy inference system are more accurate and reliable than the prediction results of the BP neural network. At the same time, it can realize real-time monitoring and has less error control. Experimental results demonstrate that in the entire production logging process of non-vertical wells, the use of a fuzzy inference system to predict fluid flow patterns can greatly save production costs while ensuring the safe operation of production equipment.

**Keywords**：production logging; oil-water two-phase flow; fuzzy inference system; non-vertical well

## 1. Introduction

The phenomenon of multiphase flow exists widely in real life and nature. In the petroleum industry, multiphase flow refers to a mixture of more than two different phases, including gas-oil, gas-water, oil-water or oil-gas-water three-phase flow. Since the 1980s, the exploitation of offshore oil has received attention from the domestic and international petroleum industries. Due to the complexity of offshore oil production, non-vertical well drilling technology can improve oil recovery while drastically reducing oil field development costs. Therefore, non-vertical well drilling technology is used extensively at home and abroad. However, compared with vertical wells, non-vertical wells have gravity differences, and fluid flow patterns in the wellbore of non-vertical wells will change dramatically. The difference between the two mining technologies results in very different equipment, instruments and mining methods.

* Corresponding author.
Email:wuyuyan2019@outlook.com, Tel:15071143168, Yangtze University

In recent decades, scholars at home and abroad have adopted various methods to predict the flow pattern of multiphase flow, and have made certain progress. In 1996, Nadler et al. divided the flow pattern diagram into: stratified flow (ST), stratified flow with mixing at the interface (ST&MI), dispersion of oil in water and water (DO/W & W), oil in water emulsion (O /W), dispersions of water in oil and oil in water (DW/O & O/W) and water in oil emulsion (W/O), dispersion of water in oil and water (DW/O & W) [1]. In 1999, Angeli et al. divided the flow patterns of oil-water two-phase flow in horizontal wells into: wave-like stratified flow (SW), mixed-interface wave-like stratified flow (SWD), three-layer flow (3L), stratified mixed flow (SM) ), completely dispersed or mixed flow (M) [2]. In 2010, Reza Ettehadi Osgouei et al. used a second-order discriminant analysis method to develop a gas-liquid flow pattern model for the apparent velocities of the liquid and gas phases in the horizontal annular geometry, and derived the corresponding functional expression [3]. At the same time, the corresponding flow pattern model was reported to the flow pattern models of Beggs & Brill [4] and Taitel & Dukler [5]. The comparison results are proved that the new gas-liquid flow pattern can more accurately predict the two-phase flow in the horizontal annulus state.

Owing to the continuous updating of experimental equipment and the complexity of the oil-water two-phase flow law, previous studies have not been able to obtain accurate flow pattern predictions. The flow pattern is directly observed and measured by visual inspection, with the help of related instruments and technology, and the corresponding empirical chart or formula is obtained. There are errors such as subjectivity and blindness. Lacking of theoretical analysis of additional factors affecting the flow pattern of multiphase flow, the identification results or comparison results cannot meet the actual needs of production logging.

With the rapid development of machine learning, the spirit of machine learning algorithms has been extensively employed in various industries. In 2016, Al-Naser et al. used Artificial Neural Network (ANN) for fluid flow pattern recognition [6]. They only used three dimensionless input variables, namely liquid Reynolds number, gas Reynolds number and pressure drop multiplier, to identify the two-phase flow (gas/liquid) flow pattern in the horizontal pipe. Build a highly accurate (over 97%) classification fluid flow pattern model that is widely used for flow pattern recognition. It can be seen that, in the recognition and prediction of multiphase flow patterns, the use of machine learning algorithms is more technically capable of intelligently identifying multiphase flow patterns than traditional methods, and it is also more accurate; it can improve the oil phase economically output, while saving production costs. However, errors in the number of samples and measurement samples may result in erroneous conclusions.

Aiming at the number of samples and the errors in the measurement samples, more and more researchers have applied the fuzzy inference system (FIS) to the flow pattern prediction of multiphase flow. In 1987, Barnea et al. first applied the Fuzzy Inference System (FIS) to predict the flow pattern of gas-liquid two-phase flow [7]. In 2015, Florentina Popa et al. proposed a novel method to identify the gas-liquid two-phase flow pattern and estimate the pressure drop [8]. They used a data-driven classification model based on FIS: superficial velocities of the gas and liquid, pipe inclination angle were used as input variables, and the flow patterns in two-phase flow were used as the output. It is useful to noting that in some areas, the fuzzy system cannot correctly predict the flow pattern due to experimental errors caused by the discretization of FIS.

The main objective of this work is to use FIS for oil-water flow patterns identification. In checking to make sure the accuracy of the fuzzy system, under the same experimental conditions, the results of the FIS and the BP neural network were compared. Through comparison, we found that FIS has higher accuracy and is more in line with actual production logging requirements while saving costs.

The paper is organized as follows:Section2 describes experimental work, Section3 explains the experimental process of the BP neural network,Section4 explains the experimental process of the fuzzy inference system,Section5 presents and discusses the results of differences between BP and FIS prediction data in this work and finally is a conclusion.

## 2. Experimental work

To conduct the experimental work, Multiphase Flow Laboratory of the Key Laboratory of Exploration Technologies for Oil and Gas Resources of Ministry of Education of Yangtze University was used to do a two-phase (oil-water) flow experiment. The schematic diagram of the laboratory's equipment is shown in Fig. 1. The length of the shaft is 12m. There is a 1m-long stainless steel pipe section at the bottom and top of the simulated wellbore for observation and measurement. The main part consists of a 10m-long glass round pipe. After the oil and water are fully mixed, there is a connecting pipe between the two wellbores. This connecting pipe is installed There is a controllable electric ball valve to control the connection or disconnection of the two wellbores, so as to conduct multiphase flow experiments repeatedly and continuously. Liquid storage tank part includes an oil storage tank, a water storage tank and an oil-water separation tank. The oil storage tank and the water storage tank store the white oil and tap water employed in the experiment. The separation tank is used to separate the oil and water. The pressure source mainly includes the water pump and the oil pump. By absorbing adverse liquid from the water storage tank and the oil storage tank, the pump into the surge tank on the derrick. The gate division includes flow control valves and fluid flow control valves, which are utilized to adjust the flow of water and oil.

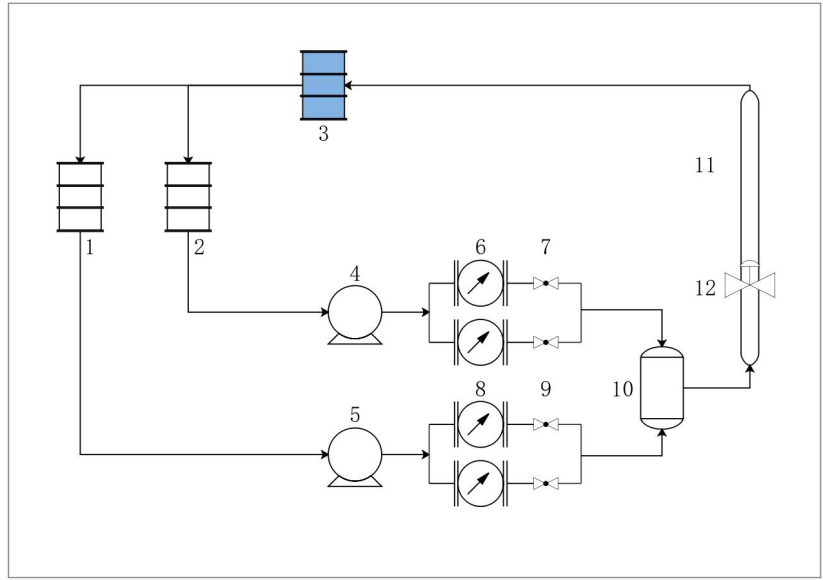

1-water storage tank, 2-oil storage tank, 3-oil-water separation tank, 4-oil pump, 5-water pump, 6,8-flow meter, 7,9-positioning control valve, 10-oil-water mixer, 11-simulation Wellbore, 12-well deviation regulator

Fig. 1 Schematic of experimental setup.

The experiment was all carried out in a simulated pipeline with an inner diameter of 159mm. During the experiment, the original data was accounted for and photographs and videos were taken. Figure 2 is a schematic diagram of the oil-water two-phase flow pattern and the high camera recordings: smooth stratified flow (SS ), stratified flow with mixing at the interface (ST&MI), water in oil emulsion (W/O), dispersions of water in oil and oil in water (DW/O & O/W), dispersion of oil in water and water ( DO/W & W), respectively.

The experimental conditions are natural temperature (20°C) and normal pressure. When the wellbore is horizontal, the inclination angle is 90°. The medium we use is No. 10 industrial white oil and tap water. White oil is a Newtonian fluid under normal conditions, with a density of 0.8263g/cm3 and a viscosity of 2.92mPa·s. The tap water has a density of 0.9884g/cm3 and a viscosity of 1.16mPa·s. The design points for the total flow of the experiment are 100 m3/d, 300 m3/d, 600 m3/d, the water content varies from 0 to 100%, and the design points are 20%, 40%, 60%, 80%, 90%, inclined The angles are 0°, 60°, 85°, 90°.

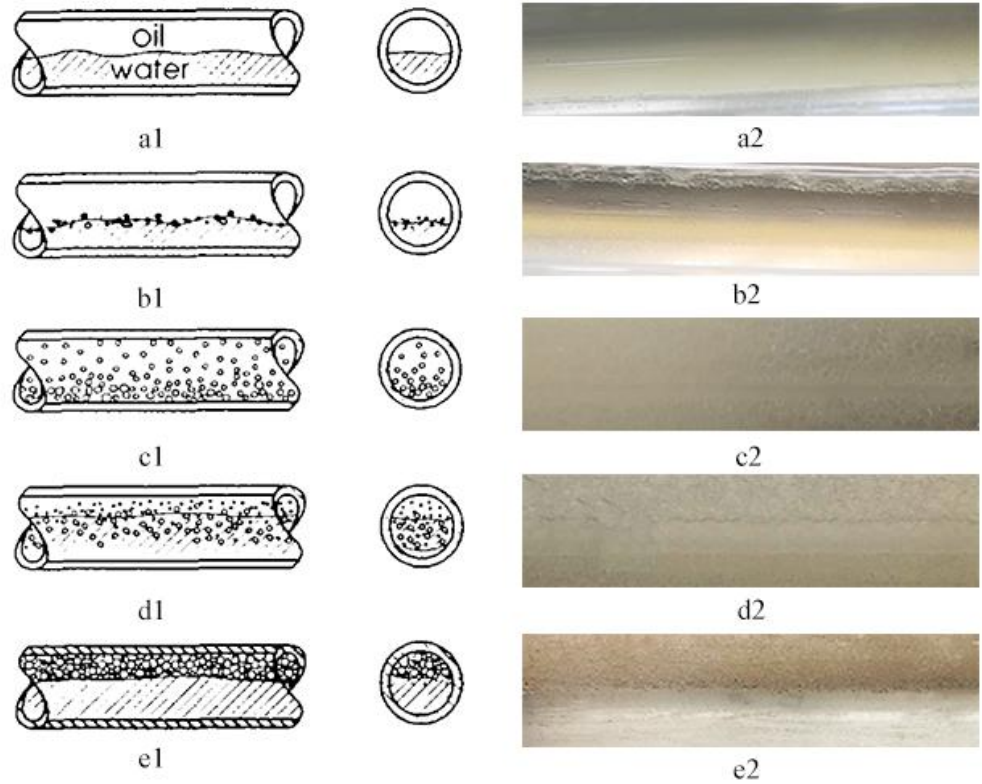

Fig. 2 Schematic of oil-water flow patterns (left) and photographed diagram (right). a1, b1, c1, d1, e1 are the schematic of SS, ST&MI, W/O, D W/O & O/W, D O/W & W, respectively (schematic source: literature [1]). a2, b2, c2, d2, e2 are the high camera recordings of SS, ST&MI, W/O, D W/O & O/W, D O/W & W, respectively.

## 3. BP (Back-propagation) neural network

BP neural network is a multi-layer feedforward network trained by error back propagation. Its basic idea is the gradient descent method, which uses gradient search technology to minimize the error and mean square error between the actual output value and the expected output value of the network. The BP neural network consists of an input layer, a hidden layer and an output layer. The model diagram is shown in Fig. 3.

For statistical convenience, we combine smooth stratified flow (SS) and stratified flow with mixing at the interface (ST&MI) into separated flow (ST). Since the neural network cannot handle non-numerical output, the four flow patterns are replaced by target values, as showed in Table 1. The prediction results are achieved by BP neural network training 42 sets of sample data are shown in Fig. 4. The red solid line curve is the predicted value, and the blue point is the actual value. We can see that the predicted value and the actual value very close. The training process is shown in Fig. 5. When the number of training reaches 300 time, the training effect is the best.

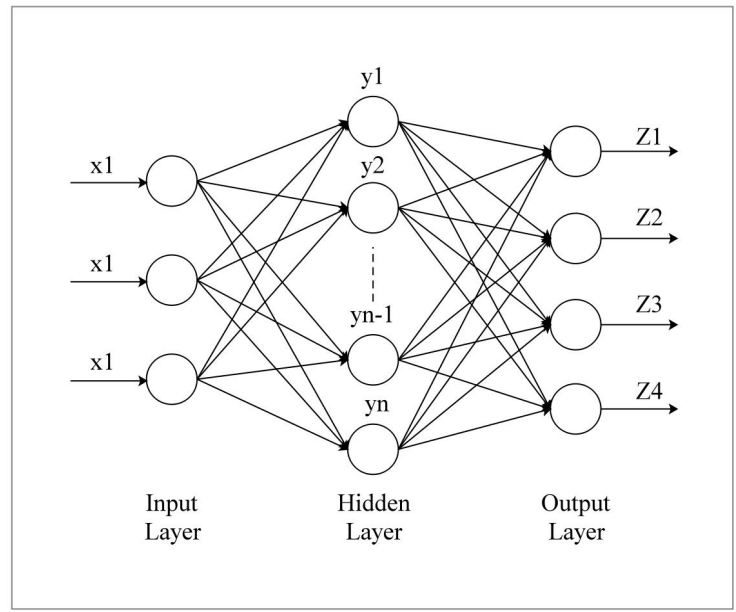

Fig. 3 Three-layer BP neural network model diagram.

### 3.1 Implementation steps

When using the artificial neural network to predict the oil-water two-phase flow pattern, first select the feature quantity and the number of training set samples to make the processing results of the training set samples reach the optimal effect, and then process other samples to be processed. This research chooses the tilt angle. The total flow rate and the water content are the network input features, and the number of training samples selects a total of 60 experimental data points. The following is the entire implementation process:

(1) Generate a BP network, the number of hidden layers is for two, the number of iterations is 300, the learning rate is 0.05, and training accuracy required is 0.00001.

(2) proceed with network training. Choosing the appropriate training function and training method. This research chooses the S-tangent function tansig as the activation function of the hidden layer neurons. The network training function is trainrp, and the network performance function is mse.

(3) The data are made available on the network, and the predicted value is obtained.

For statistical convenience, we combine smooth stratified flow (SS) and mixed interface stratified flow (ST&MI) into separated flow (ST). Since the neural network cannot handle non-numerical output, the four flow patterns are replaced by target values, as shown in Table 1. Figure 4 is the prediction result obtained by BP neural network training 42 sets of sample data. The red solid line curve is the predicted value, and the blue point is the actual value. We can see that the predicted value and the actual value are very close. Fig. 5 is the training process. When the number of training reaches 300 times, the training effect is the best.

Table 1 Numerical representation of flow patterns.

| Flow pattern | Numerical value |
|---|---|
| W/O | 1 |
| ST | 2 |
| D O/W & W | 3 |
| D W/O & O/W | 4 |

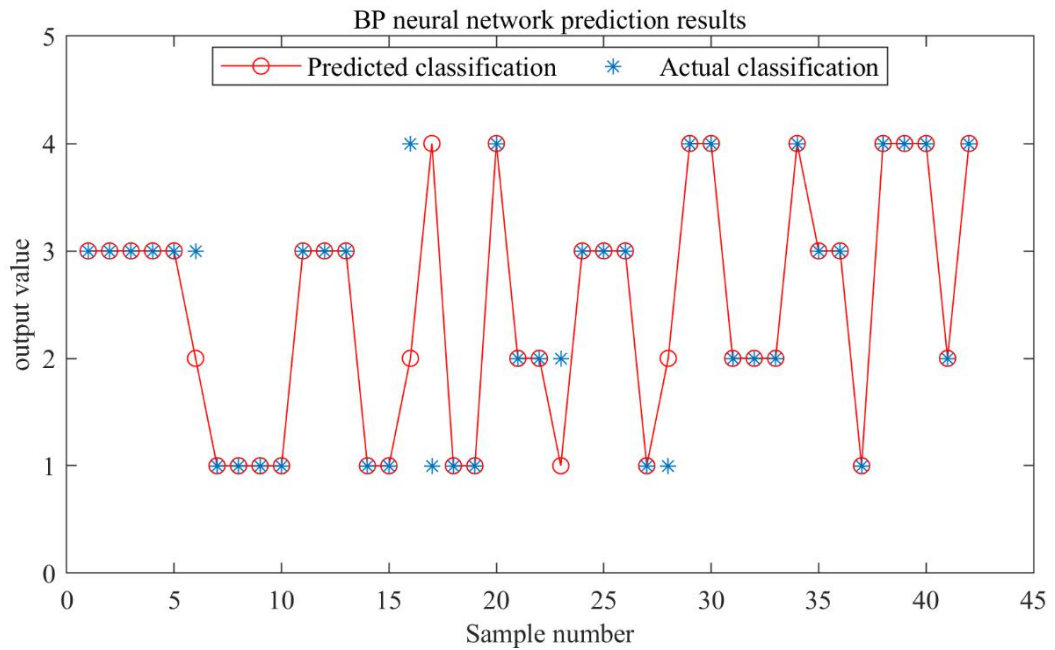

Fig. 4 BP neural network training results.

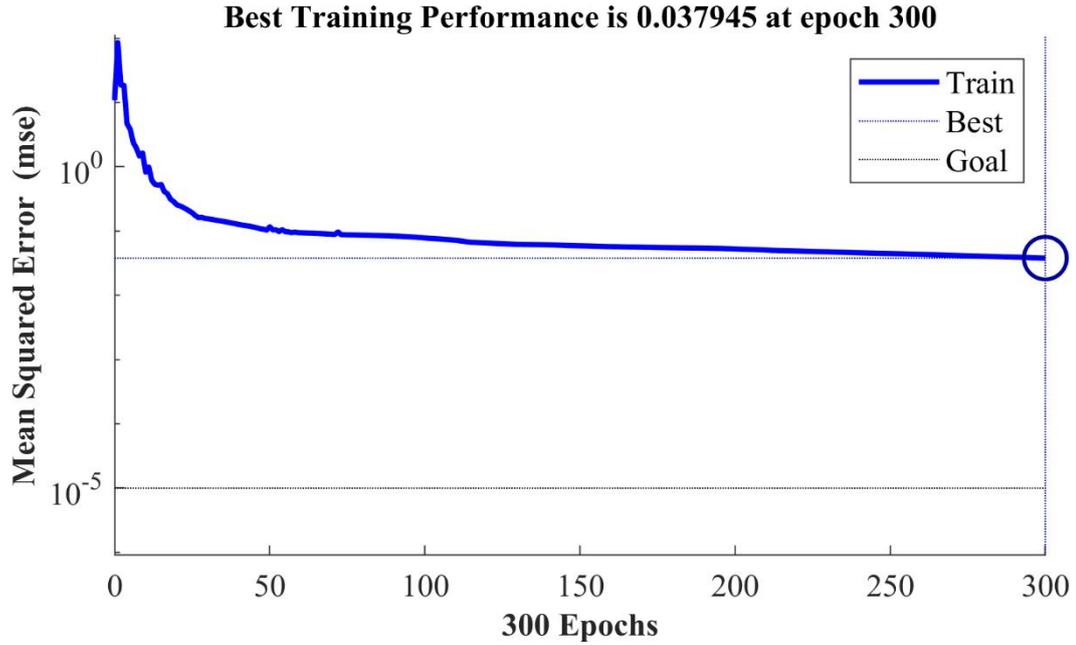

Fig. 5 BP neural network training process.

## 4. Fuzzy inference system

Fuzzy inference system (FIS) is a modeling method based on fuzzy set theory. It can perform inference on fuzzy objects that are difficult to express in mathematics by establishing fuzzy set theory, if-then rules and fuzzy inference methods [9]. FIS is essentially a mathematical method based on traditional logical reasoning (0 and 1). It can imitate the fuzzy thinking of the human brain, and carry out reasonable reasoning, perceiving and predicting fuzzy objects. In 1965, Zadeh of the United States was released famous paper "Fuzzy Sets", which established the mathematical foundation for the establishment of fuzzy set theory [10]. In 1974, British E.H. Mandani applied fuzzy logic and fuzzy reasoning to real life for the first time [11]. He realized the world's leading experimental steam engine control and achieved better results than traditional digital control algorithms. Since then, fuzzy theory has become a hot topic and has been extensively used in various neighborhoods, such as the petroleum industry and earthquake prediction.

FIS modeling comprises two parts: fuzzy system structure identification and parameter identification. At present, the division of input/output space and the mapping relationship of input/output space are called fuzzy system structure identification. Its main task is to select input and output variables, select fuzzy inference methods, and determine the correlation of each input/output variable. The number of fuzzy subsets and design a set of (if-then) fuzzy rules. The use of a certain criterion (for example, the least square criterion) to identify all parameters in the model is called parameter identification, and its principal task is to select the appropriate parametritis MF family [12].

**4.1 The realization process of FIS**

FIS flow chart is shown in Fig. 6. The following is the corresponding process:

(1) Input variables. Fuzzification interface is made on the variable, according to the membership function, the accurate value corresponds to the membership degree in the fuzzy set (a value between 0 and 1).

(2) The establishment of fuzzy rules. Use a precise logical way to combine input data to construct a set of rules and generate corresponding output results. In FIS, all the rules will be made at the same time, and each rule will draw the strength of the conclusion, that is, the degree of membership of the output fuzzy set.

(3) Determine the fuzzy reasoning method. The Mamdani method was used in this study. This method can describe the opinions of experts in a more direct and human way.

(4) Defuzzification. When converting the membership degree of the output fuzzy set into an accurate value, the step of diversification is needed. Through the inference method, a single value that best represents this fuzzy set is obtained to defuzzify. The defuzzification method used in this study is centroid.

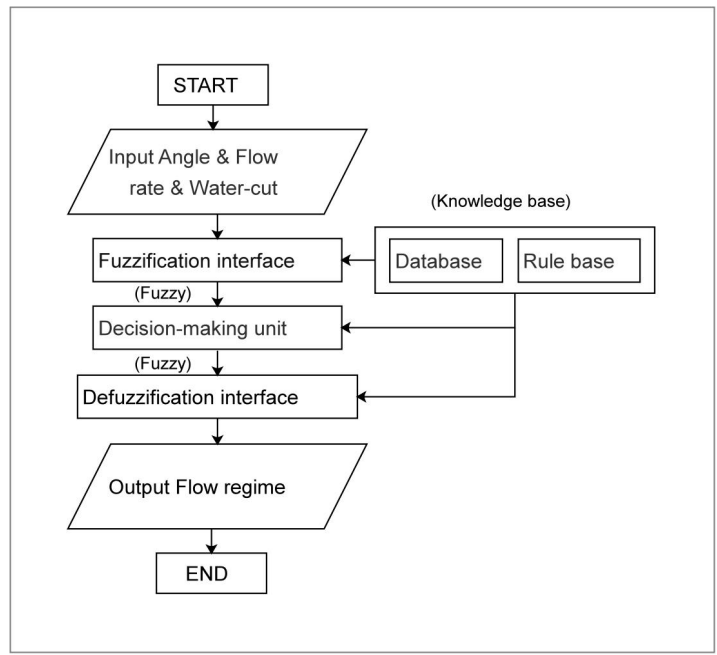

Fig. 6 FIS flow chart.

## 4.2 Membership functions for the input variables

The choice of membership function for discrete input variables is given in the expert system of system behavior. The influencing factors of oil-water two-phase flow pattern mainly include: oil and water viscosity and density, pipe diameter, pipe angle, phase holdup and other factors. During the experiment, it was found that the angle of the pipeline, the total oil-water two-phase flow, and the water cut have the most influence on the flow pattern. Therefore, these three are used as input variables to capture changes in flow patterns. The inclination angle of the pipe ranges from 0° to +90°. We divide this range into three orders of magnitude, which is semantic sets composed of the following three subsets (Fig. 7):

Positive Small(PS)

Positive(P)

Positive Large(PL)

The total flow range is 100m3/d~600m3/d. We divide this range into three orders of magnitude, which is semantic sets composed of the following three subsets (Fig. 8):

M(Middle)

H(High)

VH(Very High)

The water content ranges from 0 to 1 (dimensionless). This range is divided into five orders of magnitude, which is semantic sets composed of the following five subsets (Fig. 9):

VL(Very Low)

L(Low)

ML(Middle Low)

M(Middle)

H(High)

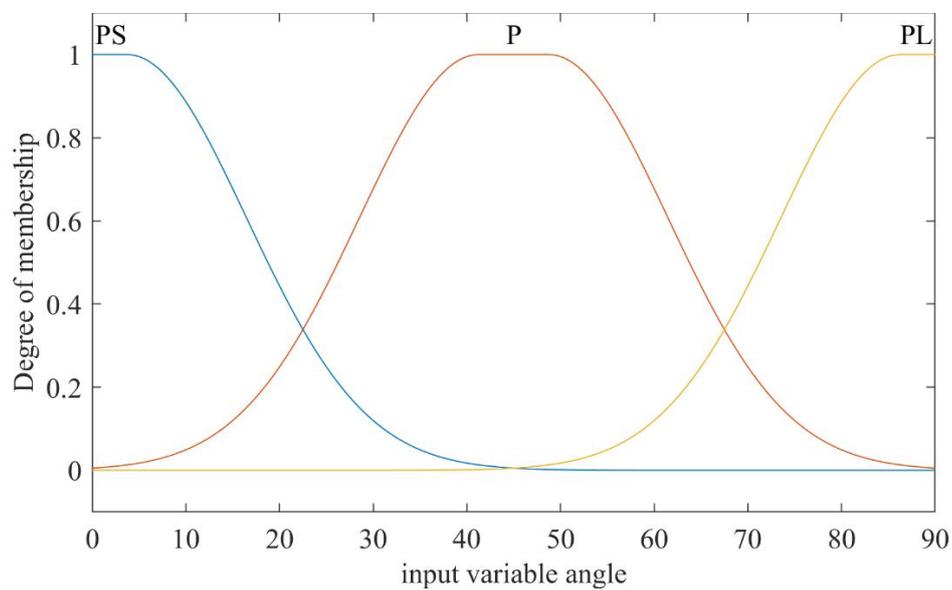

Fig. 7 The membership function of the input variable angle θ(°).

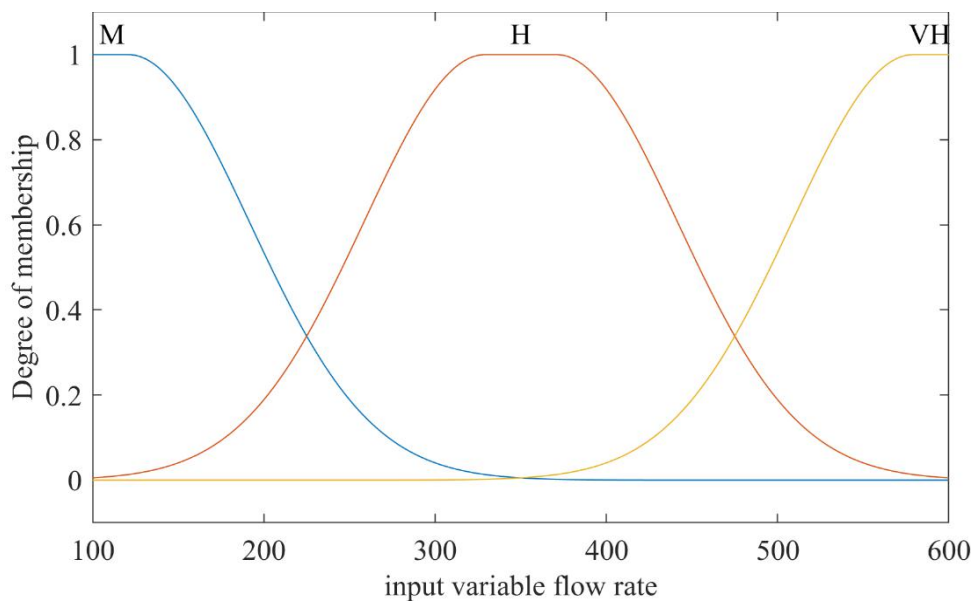

Fig. 8 The membership function of the input variable total flow (m³/d).

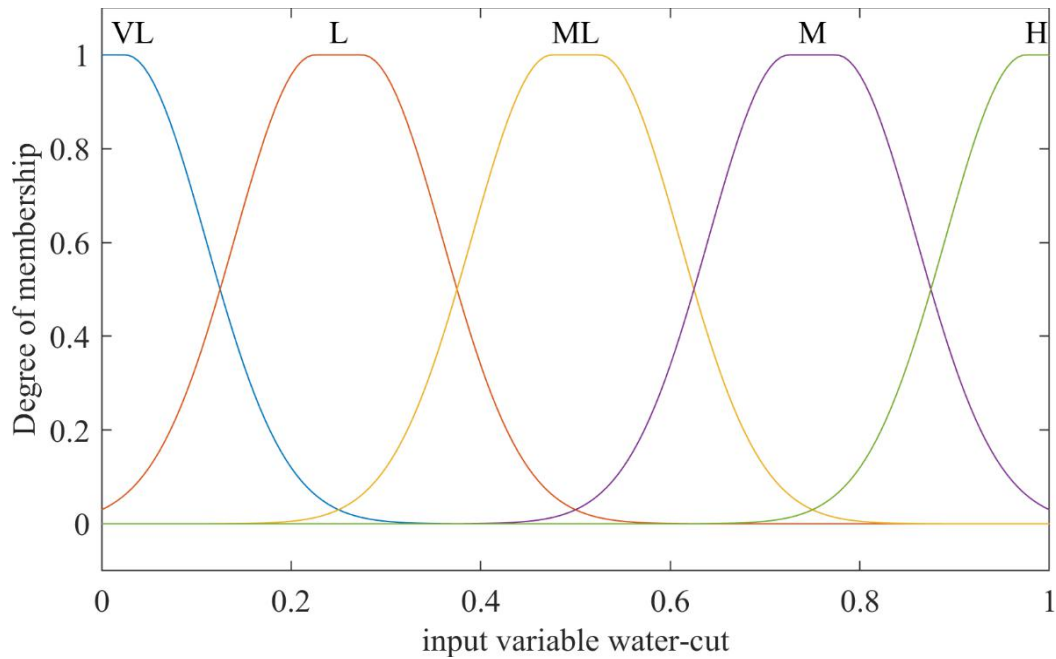

Fig. 9 The membership function of the input variable moisture content.

### 4.3 Membership functions for the output flow patterns

In this study, the oil-water two-phase flow pattern is divided into two types: separation flow and dispersed flow. Separated flows include smooth stratified flow (SS) and stratified flow with mixing at the interface (ST&MI). The dispersed flow includes dispersion of oil in water and water (D O/W & W), dispersions of water in oil and oil in water (D W/O & O/W), and water in oil emulsion (W/O).

In order to more conveniently describe the flow pattern and determine the specific position of the flow pattern, for a given input variable, the following semantic values are used to describe the position of the flow pattern (Fig. 10):

AWAY: When the flow pattern is placed in the current position, determine the unpredicted target flow pattern of the flow pattern.

FAR: When the flow pattern is at the current position, it is determined that the flow pattern is close to the expected target flow pattern.

BORDER: When the flow pattern is confirmed in the current position, it is determined that the flow pattern is very close to the expected target flow pattern (adjacent flow pattern).

CLOSE: When the flow pattern is acceptable in the current position, it is determined that the flow pattern may be the expected target flow pattern.

IN: When the flow pattern is right on the contemporary position, confirm that the flow pattern is consistent with the predicted target flow pattern.

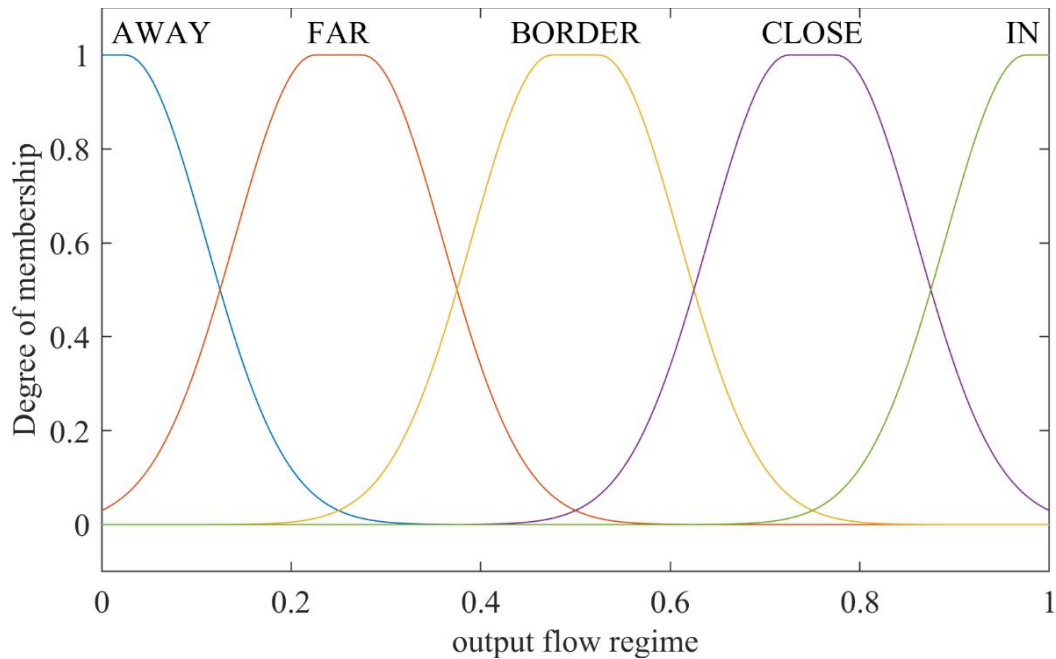

Fig. 10 Membership function of output variable flow patterns.

**4.4 Build a rule base**

This research passes the Mamdani reasoning method and establishes 20 rules in total. After defuzzification, the center of gravity method is used to aggregate all fuzzy conclusions, and the final output single value is used as the prediction result. Figure 11 is a schematic diagram of the rules. When the angle is 45°, the total flow rate is 350m3/d, and the water content is 0.5, each rule is calculated at the same time. After polymerization by the center of gravity method, the maximum value is $\Phi_{W/O}=0.99$, so The predicted flow pattern is W/O.

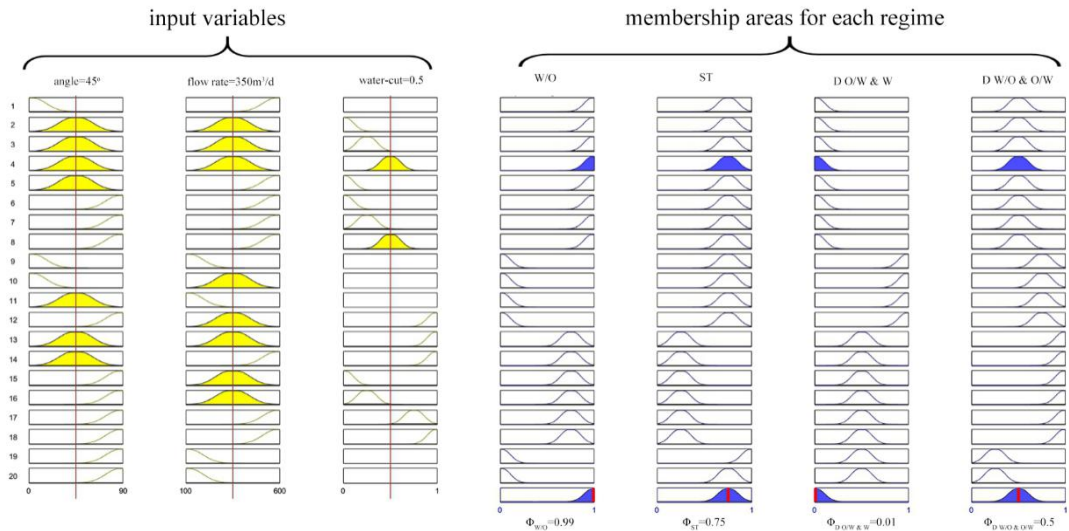

Fig. 11 Schematic of the rule base

## 5. Results and discussion

In order to provide for the oil-water two-phase flow pattern more accurately, this paper adopts FIS to reduce the interference of human subjective factors and make the prediction results more in line with actual production logging requirements. In the course of the experiment, through the analysis of the flow pattern diagrams of 0° (vertical upward), 60° (inclined upward), 85° (inclined upward), and 90° (horizontal) in Fig. 12, it was found that the oil-water two-phase flow pattern was affected. The three main factors are the angle of inclination, flow rate and water content. The angle of inclination, total flow and water cut are used as input variables to explore the changes between flow patterns. Studies have demonstrated that the results of the inference system are basically consistent with the experimental results.

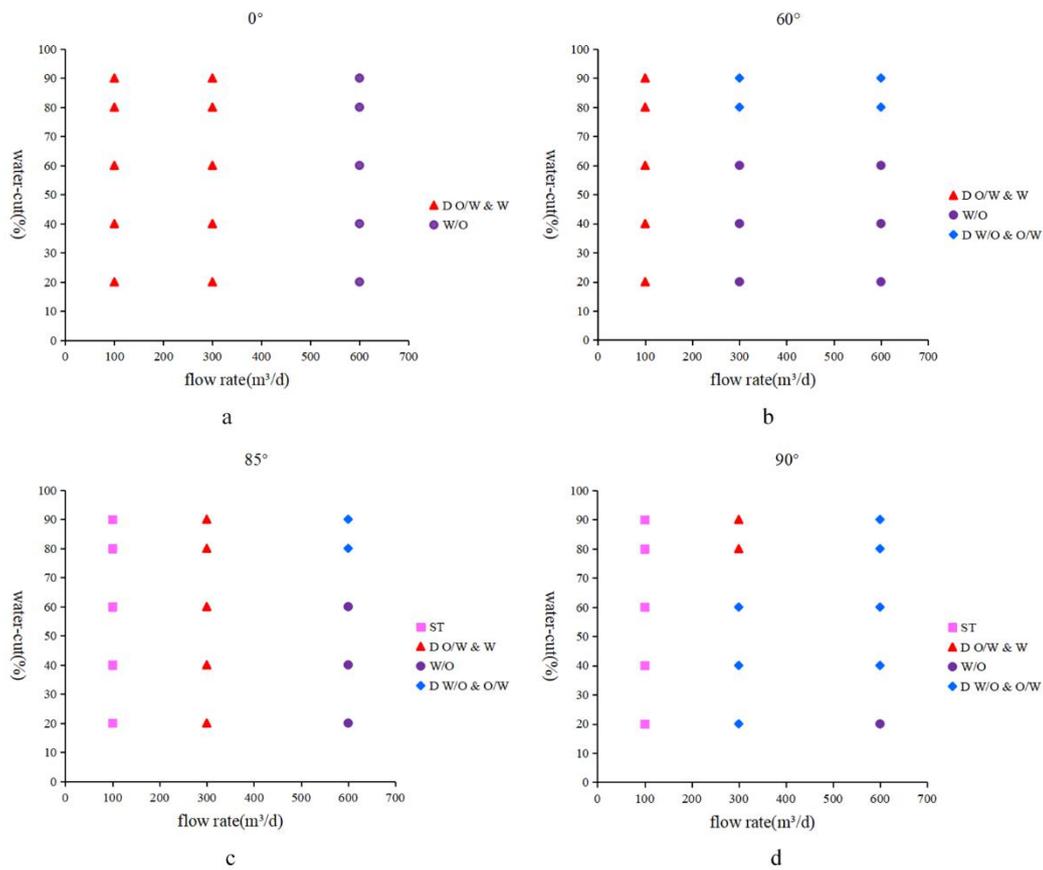

Fig. 12 a is the flow pattern when vertical upward flow (0°), b is the flow pattern when the inclination angle is inclined upward flow (60°), c is the flow pattern when the inclined upward flow (85°), d is the flow pattern when the inclination angle is horizontal flow (90°).

**5.1 Comparison of FIS and BP neural network**

Table 2 is the result of the BP neural network's prediction of the test sample points, and Table 3 is the result of the FIS's prediction of the test sample points. It can be seen from these two tables that the error value of the BP neural network is relatively large, while the error value of the FIS is relatively small.

Table 2 Flow pattern prediction results of BP test points.

| Angle (°) | Flow rate (m³/d) | Water-cut (%) | Actual pattern | Output pattern |
|---|---|---|---|---|
| 0 | 100 | 80 | D O/W & W | D O/W & W |
| 0 | 100 | 90 | D O/W & W | D O/W & W |
| **0** | **300** | **20** | **D O/W & W** | **ST** |
| **0** | **300** | **40** | **D O/W & W** | **ST** |
| 0 | 600 | 20 | W/O | W/O |
| **60** | **100** | **20** | **D O/W & W** | **D W/O & O/W** |
| 60 | 100 | 40 | D O/W & W | D O/W & W |
| 60 | 300 | 60 | W/O | W/O |
| **60** | **300** | **80** | **D W/O & O/W** | **ST** |
| 60 | 600 | 90 | D W/O & O/W | D W/O & O/W |
| 85 | 100 | 20 | ST | ST |
| 85 | 100 | 90 | ST | ST |
| 85 | 300 | 80 | D O/W & W | D O/W & W |
| 85 | 300 | 90 | D O/W & W | D O/W & W |
| **85** | **600** | **60** | **W/O** | **D O/W & W** |
| 90 | 100 | 80 | ST | ST |
| 90 | 300 | 40 | D W/O & O/W | D W/O & O/W |
| 90 | 600 | 80 | D W/O & O/W | D W/O & O/W |

Table 3 Flow pattern prediction results of FIS test points.

| Angle (°) | Flow rate (m$^3$/d) | Water-cut (%) | Actual pattern | Output pattern |
|---|---|---|---|---|
| 0 | 100 | 80 | D O/W & W | D O/W & W |
| 0 | 100 | 90 | D O/W & W | D O/W & W |
| 0 | 300 | 20 | D O/W & W | D O/W & W |
| 0 | 300 | 40 | D O/W & W | D O/W & W |
| 0 | 600 | 20 | W/O | W/O |
| 60 | 100 | 20 | D O/W & W | D O/W & W |
| 60 | 100 | 40 | D O/W & W | D O/W & W |
| 60 | 300 | 60 | W/O | W/O |
| 60 | 300 | 80 | D W/O & O/W | D W/O & O/W |
| 60 | 600 | 90 | D W/O & O/W | D W/O & O/W |
| 85 | 100 | 20 | ST | ST |
| 85 | 100 | 90 | ST | ST |
| 85 | 300 | 80 | D O/W & W | D O/W & W |
| 85 | 300 | 90 | D O/W & W | D O/W & W |
| 85 | 600 | 60 | W/O | W/O |
| 90 | 100 | 80 | ST | ST |
| 90 | 300 | 40 | D W/O & O/W | D W/O & O/W |
| **90** | **600** | **80** | **D W/O & O/W** | **ST** |

The prediction results of BP neural network have individual errors under some working conditions. When the wellbore is vertical (0°), medium flow and low water cut, DO/W & W is predicted as ST. When the inclination angle is uphill (60°), low flow rate and low water content, D O/W & W is predicted as D W/O & O/W. When the inclination angle is uphill (60°), medium flow and high water content, predict D W/O & O/W as ST. When the inclination angle is uphill (85°), high flow rate and medium water content, W/O is predicted as D O/W & W. The error value of the FIS prediction result is smaller than that of the BP neural network. When the wellbore is horizontal (90°), high flow and high water cut, predict D W/O & O/W as ST. In general, the prediction results of FIS are consistent with the experimental results, while the prediction results of the BP neural network are quite different from the experimental results.

It is concluded from the experimental results that in a highly deviated well (60°), only dispersed flow exists. When the pipeline is nearly horizontal (85°), horizontal (90°) and low flow, there is a separated flow. The prediction result of FIS satisfies the above conclusion. However, the BP neural network in highly deviated well (60°) shows a separated flow in the prediction result, indicating that there is a general error in the prediction of the flow pattern.

**5.2 Limitations of FIS**

The above analysis results show that the use of FIS can more accurately predict the oil-water two-phase flow pattern. Judging from the FIS prediction results, FIS still has errors, which may be due to the discretization of the method and the diversity and complexity of the flow pattern itself. However, it does not affect the flow pattern prediction as a whole, which is consistent with the actual production logging needs. The error of the BP neural network is relatively large. The reason for this is perhaps that the BP neural network needs to constantly adjust the weights and thresholds to predict the accurate flow pattern. It can be seen from these two that the reliability and accuracy of FIS are still relatively high, and it can be applied to the inference of uncertain objects. However, artificial neural networks still need to improve the prediction of changeable objects. Constantly adjusting the weight and system parameters can achieve a better learning effect, which consumes more time and energy.

**6. Conclusion**

This paper mainly studies the application of FIS in oil-water two-phase flow pattern. Prediction models of FIS and BP neural network in oil-water two-phase flow patterns are compared. From the comparison results, it is concluded that the accuracy of FIS's oil-water two-phase flow pattern prediction is higher than that of BP neural network, and the work efficiency is high, and the cost is kept at the same time. We use FIS to predict the oil-water two-phase flow pattern. The main advantage of this technology is to ensure that the uncertainty and complexity data of the regional flow pattern clearer and clearer. It can realize high-efficiency flow pattern recognition through fuzzy reasoning. The system saves time and labor costs. Therefore, the fuzzy inference system can realize real-time monitoring in actual production logging. Using this method to predict other uncertain variables in the actual oil field, such as solid sediment, can ensure the safe operation of production equipment while greatly saving Production cost, so as to achieve high-efficiency production requirements.

It is useful to noting that, due to the large difference between the ground experimental environment and the downhole logging environment, direct quoting of experimental results will inevitably lead to errors. Therefore, when it is applied to the multiphase flow interpretation of production logging in downhole logging environment, its calculation accuracy has yet to be verified.